\documentclass{PoS}

 \usepackage{graphics}
 \usepackage{graphicx}
 \usepackage{epsfig}
 \usepackage{longtable}
 \usepackage{latexsym}
 \usepackage{amsmath}
 \usepackage{amsthm}
 \usepackage{amsfonts}
 \usepackage{amssymb}
 \usepackage{dsfont}
 \usepackage{bm}
 \usepackage{graphics,psfrag}
 \usepackage{graphicx,psfrag}
 \usepackage{subfigure}
 \newcommand{\be}{\begin{equation}}
 \newcommand{\ee}{\end{equation}}
 \newcommand{\bey}{\begin{eqnarray}}
 \newcommand{\eey}{\end{eqnarray}}

 \newcommand{\DCI}{\ensuremath{D_\srm{CI}} }

 \newcommand{\srm}[1]{\textrm{\scriptsize{#1}}}
 
 \newcommand{\E}{\mathrm{e}}
 \newcommand{\FC}{\;,}
 
 \renewcommand{\imath}{\mathrm{i}}

\title{Some results on excited hadrons in 2-flavor QCD}

\ShortTitle{Some results on excited hadrons in 2-flavor QCD}

\author{\speaker{Georg P.~Engel}
\\
        Institut f\"ur Physik, FB Theoretische Physik, Universit\"at
Graz, A--8010 Graz, Austria\\
        E-mail: \email{georg.engel@uni-graz.at}
}
\author{C.~B.~Lang\\
        Institut f\"ur Physik, FB Theoretische Physik, Universit\"at
Graz, A--8010 Graz, Austria\\
        E-mail: \email{christian.lang@uni-graz.at}
}
\author{Markus Limmer\\
        Institut f\"ur Physik, FB Theoretische Physik, Universit\"at
Graz, A--8010 Graz, Austria\\
        E-mail: \email{markus.limmer@uni-graz.at}
}
\author{Daniel Mohler\\
        TRIUMF, 4004 Wesbrook Mall Vancouver, BC V6T 2A3, Canada \\
        E-mail: \email{mohler@triumf.ca}
}
\author{Andreas Sch\"afer\\
        Institut f\"ur Theoretische Physik, Universit\"at
Regensburg, D--93040 Regensburg, Germany\\
        E-mail: \email{andreas.schaefer@physik.uni-regensburg.de}
}
\abstract{
Results of hadron spectroscopy with two dynamical mass-degenerate 
chirally improved quarks are presented. Three ensembles with pion masses of $322(5)$, $470(4)$ and
$525(7)$ MeV, lattices of size $16^3 \times 32$, and lattice spacings close to $0.15\,$ fm are investigated. 
We discuss the possible appearance of scattering states, considering masses
and eigenvectors.  Partially quenched results in the scalar channel suggest the
presence of a 2-particle state, however, in most channels we cannot identify
them. Where available, we compare to results from quenched simulations using the same action.
}

\FullConference{The XXVIII International Symposium on Lattice Field Theory, Lattice2010\\
		June 14-19, 2010\\
		Villasimius, Italy}

\begin{document}
%%%%%%%%%%%%%%%%%%%%%%%%%%%%%%%%%%%%%%%%%%%%%%%%%%%%%%%%%%%%%%%%%%%%%%%%%%%%%%%

%%%%%%%%%%%%%%%%%%%%%%%%%%%%%%%%%%%%%%%%%%%%%%%%%%%%%%%%%%%%%%%%%%%%%%%%%%%%%%%
\section{Introduction} 
%%%%%%%%%%%%%%%%%%%%%%%%%%%%%%%%%%%%%%%%%%%%%%%%%%%%%%%%%%%%%%%%%%%%%%%%%%%%%%%
\noindent
The majority of hadronic states in the Particle Data Group's
collection are hadron excitations \cite{PDG08}. 
So far, lattice QCD provides the only known way to perform ab-initio calculations of the corresponding 
observables. This  article is another step in this enterprise.
We use the Chirally Improved (CI) Dirac operator \cite{Gattringer:2000js}, which is an approximate solution
of the Ginsparg-Wilson (GW) equation \cite{Ginsparg:1981bj}.
We present
results of ground states as well as excited states, making use of the
variational method. In addition to the
light quarks we also consider heavier valence
(strange) quarks and include strange hadrons in our analysis. 
We discuss the possible appearance of scattering states and compare to quenched results using the same action.
Preliminary results have been presented in \cite{Engel:2009cq}.
A more complete discussion of the results is found in \cite{Engel:2010my}.

%%%%%%%%%%%%%%%%%%%%%%%%%%%%%%%%%%%%%%%%%%%%%%%%%%%%%%%%%%%%%%%%%%%%%%%%%%%%%%%
\section{Simulation details}
\label{SimDetSec}
%%%%%%%%%%%%%%%%%%%%%%%%%%%%%%%%%%%%%%%%%%%%%%%%%%%%%%%%%%%%%%%%%%%%%%%%%%%%%%%
\noindent
All details of the simulation method 
are given in \cite{Gattringer:2008vj}. 
The Chirally Improved Dirac  operator ($\DCI$)
is obtained by insertion of the most general
ansatz for a Dirac operator into the GW equation and comparison of the
coefficients. 
Furthermore, we include one level of stout smearing 
as part of the
definition of  $\DCI$, 
and use the L\"uscher-Weisz gauge action. % as discussed in \cite{Gattringer:2008vj}.
We generate the dynamical
configurations with a Hybrid Monte-Carlo (HMC) algorithm.
We simulate three ensembles with lattices of size $16^3 \times 32$, for details see Table \ref{SimDetTab}. 
\begin{table*}[b]
\begin{center}
\begin{tabular}{cccccccccc}
\hline
\hline
set&	$\beta_{LW}$&$m_0$& configs 	& $a$[fm]	& $m_{\pi}$[MeV]& $m_{AWI}$[MeV] 	&$L$[fm]	&$m_{\pi}L$ \\
\hline
A&	4.70& 		 -0.050	&100 	&0.151(2)	&525(7)         & 43.0(4)		& 2.42(3)	& 6.4\\
B&	4.65& 		 -0.060	&200 	&0.150(1)	&470(4)         & 35.1(2)		& 2.40(2)	& 5.7\\
C&	4.58& 		 -0.077	&200 	&0.144(1)	&322(5)         & 15.0(4)		& 2.30(2)	& 3.7\\
\hline
\hline
\end{tabular}
\end{center}
\caption{Bare parameters of the simulation: Three ensembles (A,B,C) at different gauge couplings $\beta_{LW}$ and quark mass parameters $m_0$.
The number of configurations, lattice spacing from the static potential assuming a Sommer parameter of $r_{0,exp} = 0.48\,$ fm, the pion mass, the (non-renormalized) AWI-mass, the lattice size and the dimensionless product of the pion mass with the physical lattice size are given. 
For more details see \cite{Gattringer:2008vj}.}
\label{SimDetTab}
\end{table*}
The variational method \cite{Michael:1985ne:Luscher:1990ck} is used to extract ground and excited states. % , though they are suppressed by $\mathcal{O}(\E^{-\Delta E})$.  
Given a set of interpolators (with given quantum numbers)  the corresponding correlation matrix
is
\begin{eqnarray}
C_{ij}(t) 		&=& 	\langle 0 \vert O_i(t)  O_j^{\dag} \vert 0\rangle  = \sum_{k=1}^N \langle 0 \vert O_i \vert k\rangle \langle k \vert O_j^{\dag} \vert 0\rangle\,\E^{-t E_k}.
\label{VarMethEq}
\end{eqnarray}
In the variational
method, the idea is  to offer a basis of suitable interpolators, from which the system
chooses the linear combinations closest to the physical eigenstates $|k\rangle$.
The generalized eigenvalue equation
\begin{eqnarray}\label{evshape}
C(t)\, \vec{v}_k 	&=& 		\lambda_k(t,t_0)\, C(t_0)\, \vec{v}_k \FC \qquad \lambda_k(t,t_0) 	\propto 	\E^{-(t-t_0)\,E_k} \left( 1+\mathcal{O}(\E^{-(t-t_0)\,\Delta E_k}) \right)\ ,
\end{eqnarray}
gives the energies of the eigenstates, where $\Delta E_k$ is the distance of
$E_k$ to the closest state. 
The corresponding eigenvectors represent the linear combinations of the given
interpolators which are closest to the physical states considered.
We use two Gaussian (narrow and wide) and derivative sources.
The gaussian sources are computed using gauge-covariant Jacobi smearing, 
the derivative source are obtained by applying the covariant
difference operators on the wide source. 
Combining these quark sources we construct several interpolators in each hadron channel in
order to be able to extract excited states using the variational method. All
sources are located in a single time slice and built on configurations  which
have been  hypercubic-smeared (HYP) in the spatial directions three times.
Tables of the interpolators are found in the Appendix of \cite{Engel:2010my}.
We consider isovector-mesons, which are free of disconnected diagrams. 
We use the meson interpolator construction as described in detail in \cite{Gattringer:2008be}, 
which is similar to constructions previously used in \cite{Lacock:1996vy,Liao:2002rj,Dudek:2007wv}. 
For the construction of baryon interpolators we use only Gaussian smeared quark
sources. 

In defining the lattice scale we use a mass-dependent scheme, since we have only one ensemble at each value of the gauge coupling.
Nevertheless, we assume our path in parameter space to be close to the one in the
mass-independent scheme, and expect that the analytic form of the chiral
extrapolation should be similar, although with different expansion 
coefficients. Therefore, we perform chiral fits linear in the pion mass squared
for all results.

%%%%%%%%%%%%%%%%%%%%%%%%%%%%%%%%%%%%%%%%%%%%%%%%%%%%%%%%%%%%%%%%%%%%%%%%%%%%%%%
\section{Results}
\label{Results}
%%%%%%%%%%%%%%%%%%%%%%%%%%%%%%%%%%%%%%%%%%%%%%%%%%%%%%%%%%%%%%%%%%%%%%%%%%%%%%%
\noindent
In all plots, filled symbols denote dynamical results and open
symbols denote partially quenched results. 
The energy levels are obtained by a correlated exponential fit to the leading eigenvalues
(\ref{evshape}) in a range of $t$-values where we identify a plateau behaviour of effective
mass and/or eigenvector components.

We discuss the possible appearance of scattering states considering masses, partially quenched data and eigenvectors.
Neglecting the interactions of the hadronic bound states and finite volume effects, the energy level
$E(A,B)$ for two free hadrons reads
\begin{eqnarray}
\label{eq_scatter}
 E \left( A(\vec p),B(-\vec p) \right) &=&  \left( \sqrt{m_A^2+|\vec p|^2} + \sqrt{m_B^2+|\vec p|^2} \right) \left( 1 + \mathcal{O}(ap) \right)  \; .
\end{eqnarray}
The symbols $\times$ and $+$ in the plots represent the tentative positions of expected
energy levels of free particle scattering states according to (\ref{eq_scatter}).
The corresponding non-correlated statistical uncertainty is of the magnitude of 5 to 60
MeV. 

%%%%%%%%%%%%%%%%%%%%%%%%%%%%%%%%%%%%%%%%%%%%%%%%%%%%%%%%%%%%%%%%%%%%%%%%%%%%%%%
\subsection{The 1$^{--}$ channel: $\rho$}
\begin{figure}
\begin{minipage}[t!]{74mm}
\includegraphics[width=73mm,clip]{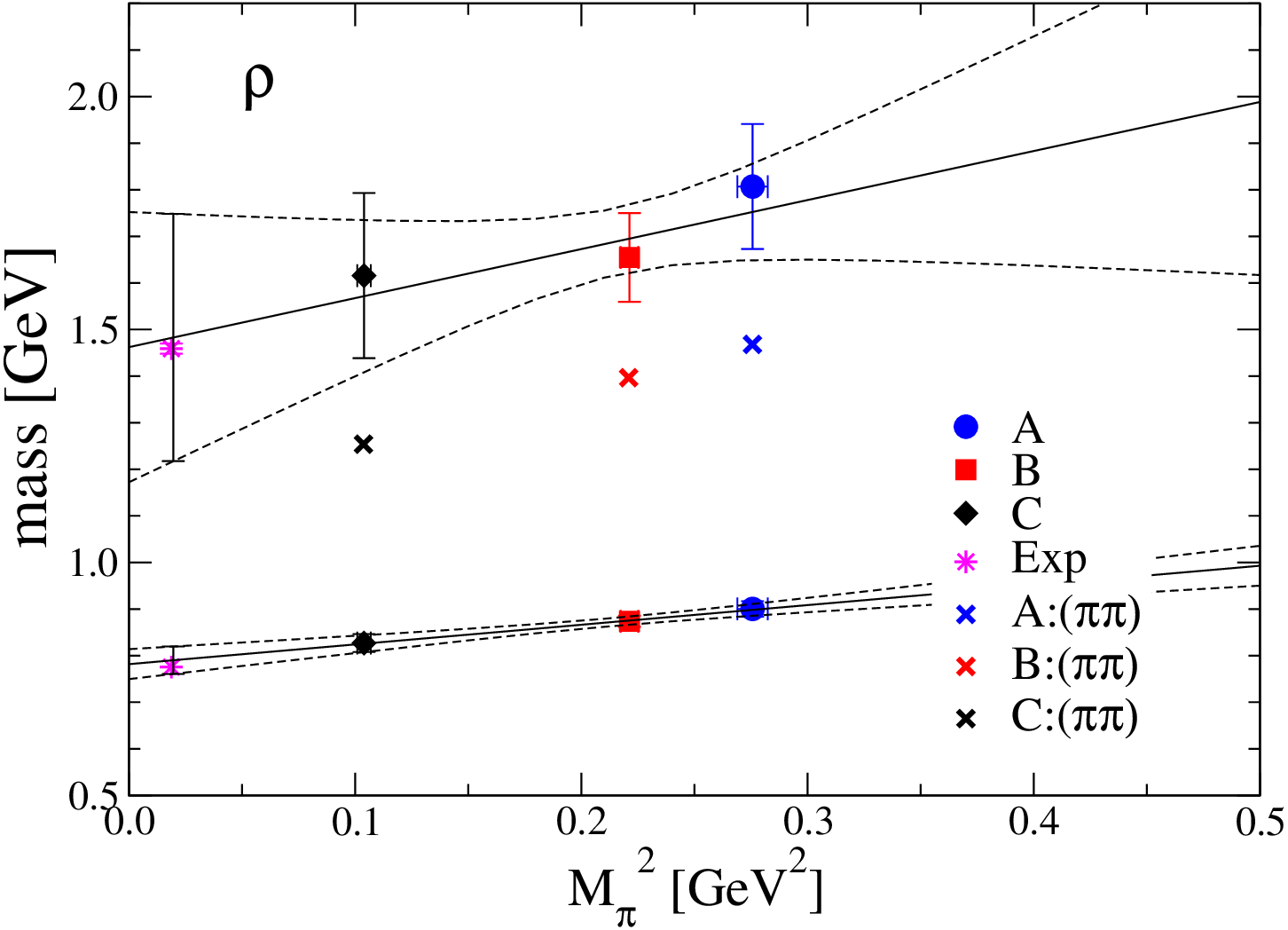}\caption{Mass plot for the 1$^{--}$ channel ($\rho$), ground state and first
excitation. 
The estimated energy level of the $P$ wave scattering state $\pi\pi$ lies
between the ground and the first excited state. 
The results suggest that the scattering state is not observed.}
\label{rho}
\end{minipage}\hfill
\begin{minipage}[t!]{74mm}
\vspace*{-0.5mm}
\includegraphics[width=73mm,clip]{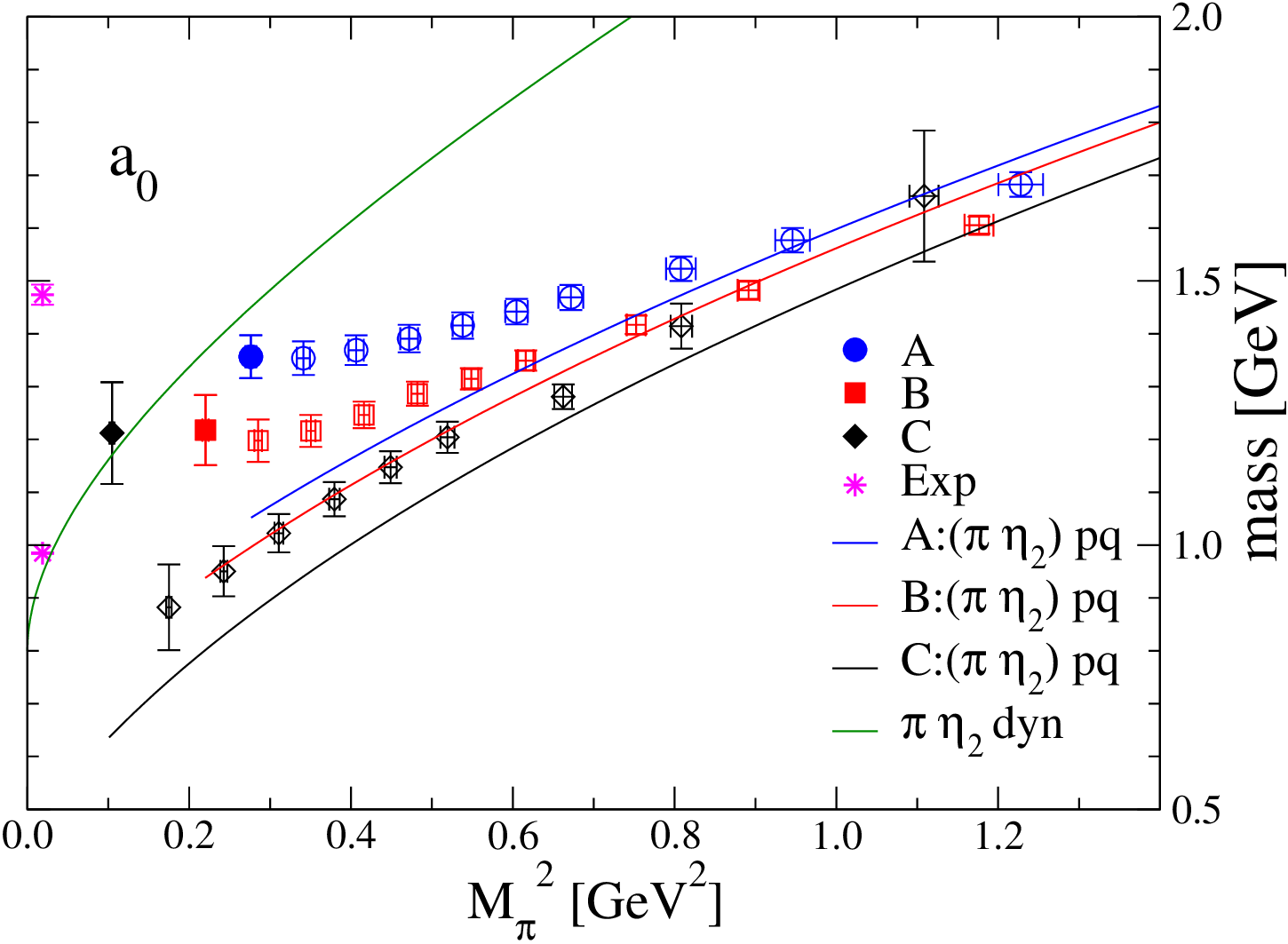}
\caption{Mass plot for the 0$^{++}$ channel ($a_0$). 
The blue, red and black curves (online
version) show a prediction of the partially quenched (``pq'') 
$\pi\eta_2$ for $m_{val}\gg m_{sea}$.
The
green curve (online version) shows an estimate of the dynamical (``dyn'')
$\pi\eta_2$ \cite{Engel:2010my}.
}
\label{a0}
\end{minipage}
\end{figure}
\noindent
We find an excellent plateau for the ground state and an excited state signal
compatible with experimental data (see Fig.~\ref{rho}).  
On the lattice, for our ensemble parameters the energy of the $P$ wave 
scattering state $\pi\pi$ would be between the $\rho$ ground state and the first excitation, 
but no such state is observed here. 
Comparing with quenched results using the same action \cite{Gattringer:2008be}
and taking into account the different Sommer parameter value used in the quenched analysis ($r_{0,exp}=0.5$ fm),
we find that the dynamical $\rho$ ground state comes out significantly lighter
than its quenched counterpart.

\subsection{The 0$^{++}$ channel: $a_0$}
\noindent
We find large effects due to partial quenching close to the dynamical point, especially at
small pion masses (see Fig.~\ref{a0}).  
Our partially quenched data are well described by
the partially quenched formulae of the scattering state \cite{Prelovsek:2004jp}, and thus are interpreted as contributions of the 2-particle state $\pi\eta_2$. 
However, at the physical point, the particle content remains unclear.
The ground state energy level in quenched simulations with the same action
\cite{Gattringer:2008be,Burch:2006dg} was extracted only at larger pion
masses, being compatible with our dynamical data of set A, extrapolating to the $a_0(1450)$ rather than to $a_0(980)$. The spectroscopy of
the light scalar channel appears to benefit from sea quarks.

%%%%%%%%%%%%%%%%%%%%%%%%%%%%%%%%%%%%%%%%%%%%%%%%%%%%%%%%%%%%%%%%%%%%%%%%%%%%%%%
\subsection{Nucleon negative parity}
\begin{figure}[tbp]
\begin{minipage}[t!]{74mm}
\includegraphics[width=74mm,clip]{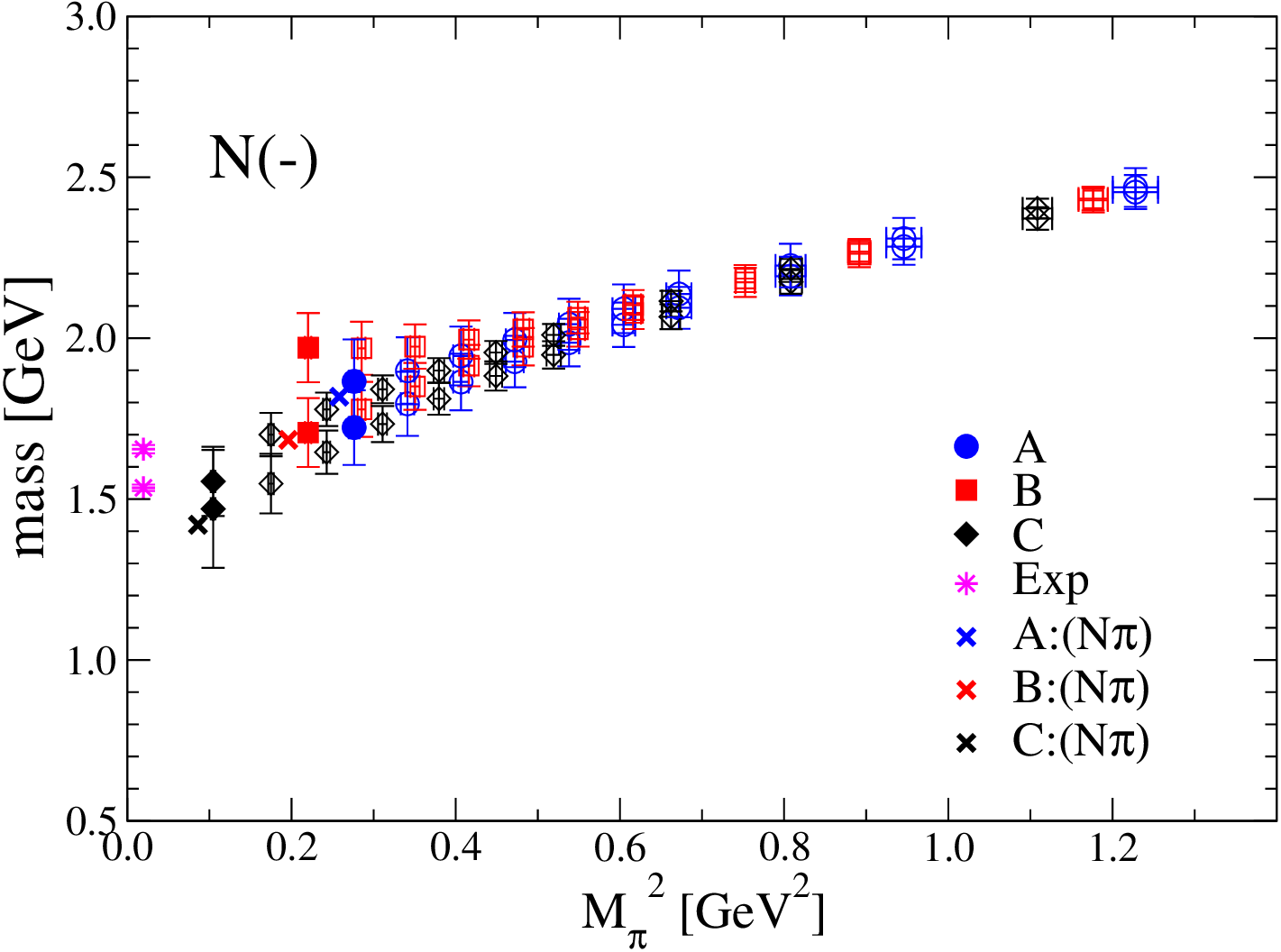}
\caption{Mass plot for the nucleon negative parity channel. 
For clarity, we display the $S$ wave scattering state $\pi N$ slightly shifted to the left.
The mass results (values below the ground state mass of N(1535)) suggest an interpretation in terms of level crossing, but the eigenvectors contradict this picture (see Fig.~5).%TODO:take care %\ref{nucleon_neg_vectors}).
Figure taken from \cite{Engel:2010my}.
} 
\label{nucleon_neg}
\end{minipage}\hfill
\begin{minipage}[t!]{74mm}
\vspace*{-5.4mm}
\includegraphics[width=74mm,height=54.3mm,clip]{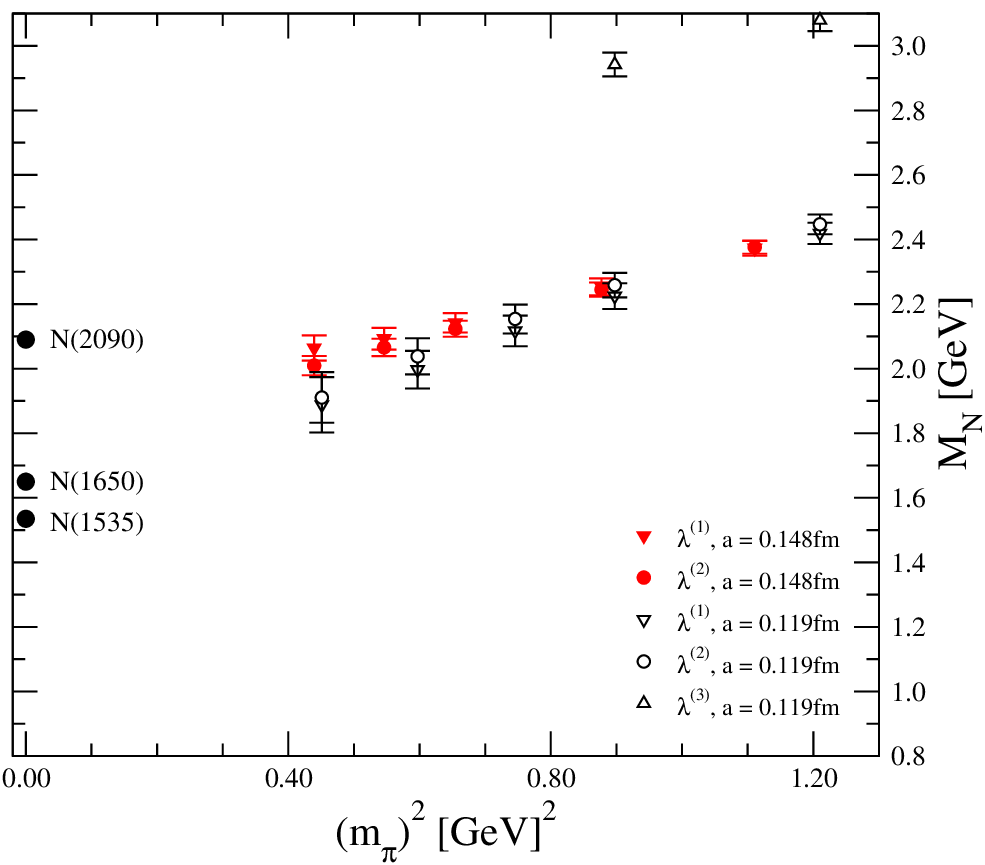}
\caption{
Mass plot for nucleon negative parity from quenched simulations using the same action.
Data is only available for pion masses larger than 400 MeV, thus the bending down of dynamical data at small pion masses cannot be compared.
Figure taken from \cite{Burch:2006cc}.
} 
\label{nucleon_neg_quenched}
\end{minipage}
\end{figure}
\noindent
The mass results suggest a large contribution of the $S$ wave scattering state $\pi N$ at small pion masses 
and thus an interpretation in terms of a level crossing for pion masses in the range of 320 to 530 MeV (see Fig.~\ref{nucleon_neg}).
Unfortunately, the bending down of the dynamical data at small pion masses cannot be compared to results from quenched simulations using the same action, 
since they are only available for pion masses larger than 400 MeV (see Fig.~\ref{nucleon_neg_quenched}).
However, considering the eigenvectors, we find that in all three ensembles the ground state is
dominated by the pseudoscalar diquark, the first excitation  by the scalar diquark interpolator (see Fig.~\ref{nucleon_neg_vectors}). 
Qualitatively, the eigenvectors of quenched simulations using the same action \cite{Burch:2006cc} show the same behavior.
Since the argument based on the eigenvectors is assumed to be more reliable than the one based on the masses, 
we may conclude that no level crossing of the lowest two states is observed for pion
masses in the range of 320 to 530 MeV and that both are mainly 1-particle states.

%%%%%%%%%%%%%%%%%%%%%%%%%%%%%%%%%%%%%%%%%%%%%%%%%%%%%%%%%%%%%%%%%%%%%%%%%%%%%%%
\subsection{Setting the strange quark mass}
\label{strangemass}
\begin{figure}[tbp]
\begin{minipage}[t!]{74mm}
\vspace*{-2mm}
\includegraphics[width=73mm,clip]{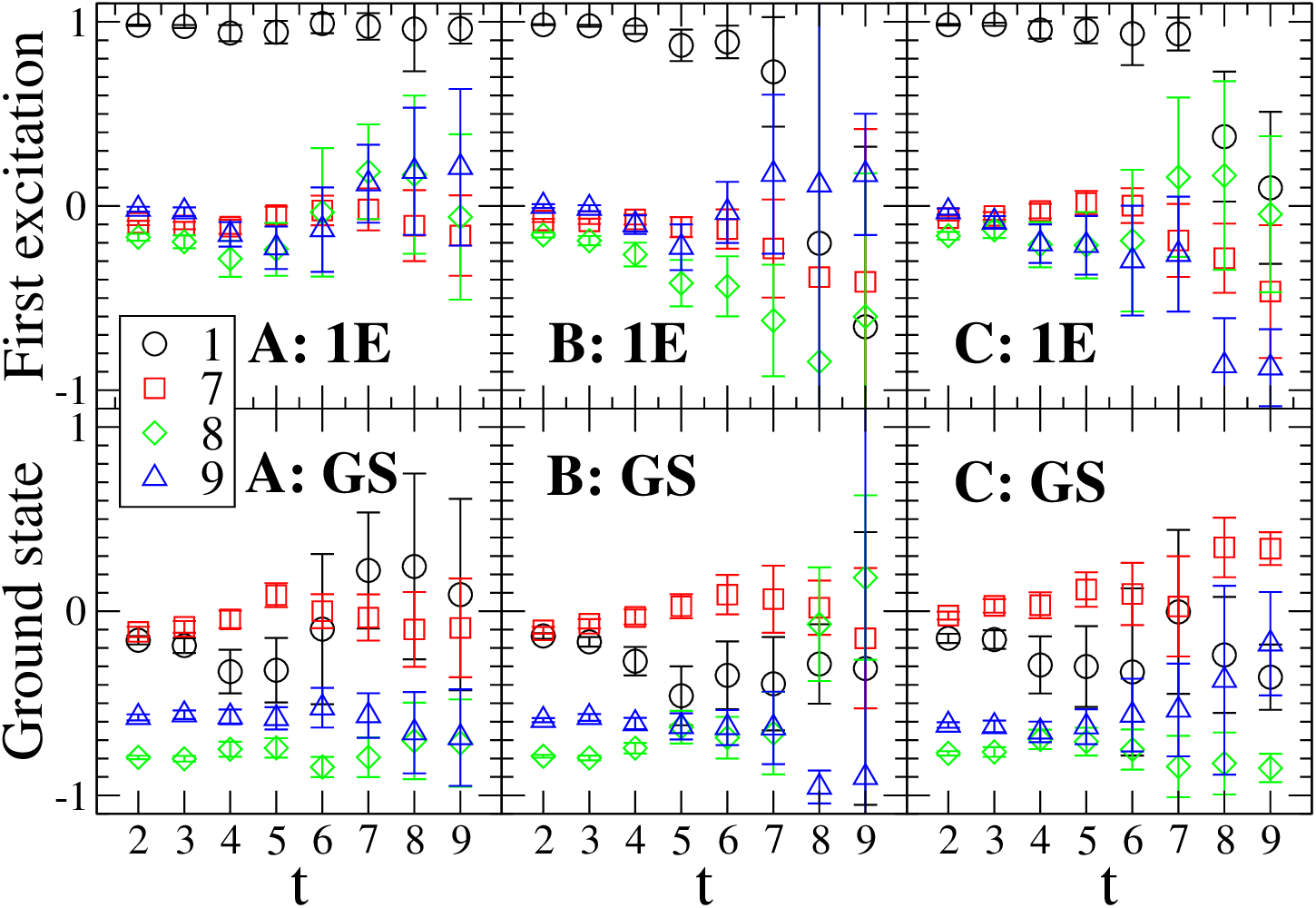}
\caption{
Eigenvectors of the nucleon negative parity channel, ground
state and first excitation at the dynamical point. 
In all three ensembles the ground state is
dominated by pseudoscalar, the first excitation
by scalar diquark (similar to the quenched results in \cite{Burch:2006cc}). % (see Table \ref{tab:baryoninterpolators1}). 
One may conclude that no level crossing of the lowest two states is observed.
Plot taken from \cite{Engel:2010my}.
} 
\label{nucleon_neg_vectors}
\end{minipage}\hfill
\begin{minipage}[t!]{74mm}
\includegraphics[width=74mm,clip]{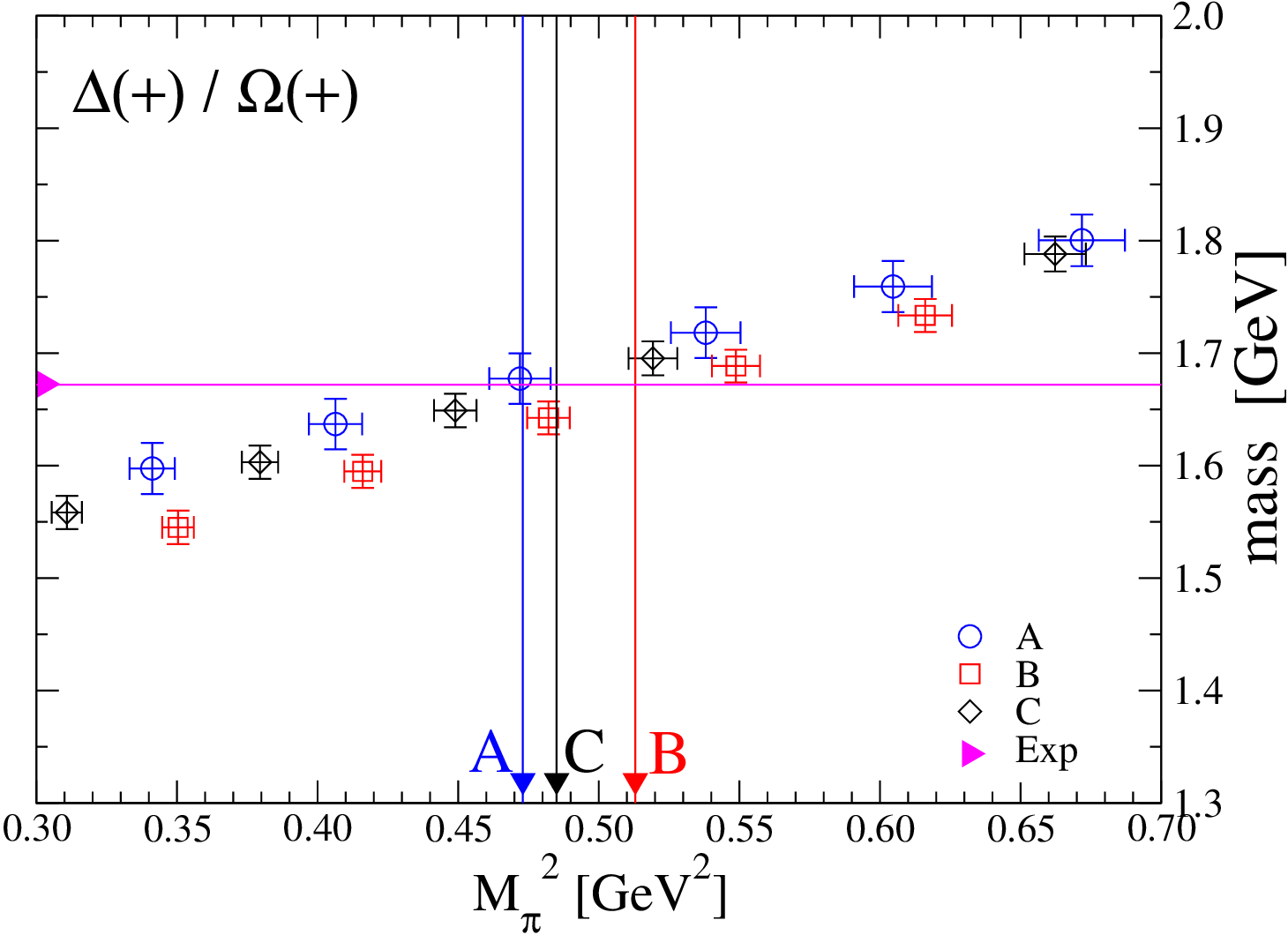}
\caption{Extracting the strange quark mass parameter by identifying a partially
quenched $\Delta$ with $\Omega$(1670), represented by the magenta (online version) horizontal line.
Crossing this line with the partially quenched $\Delta$ mass curves defines the bare strange quark
mass parameter of A, B and C, illustrated by the three vertical lines \cite{Engel:2010my}.
}
\label{omega}
\end{minipage}
\end{figure}
\begin{figure}[tbp]
\begin{minipage}[t!]{74mm}
\includegraphics[width=73mm,clip]{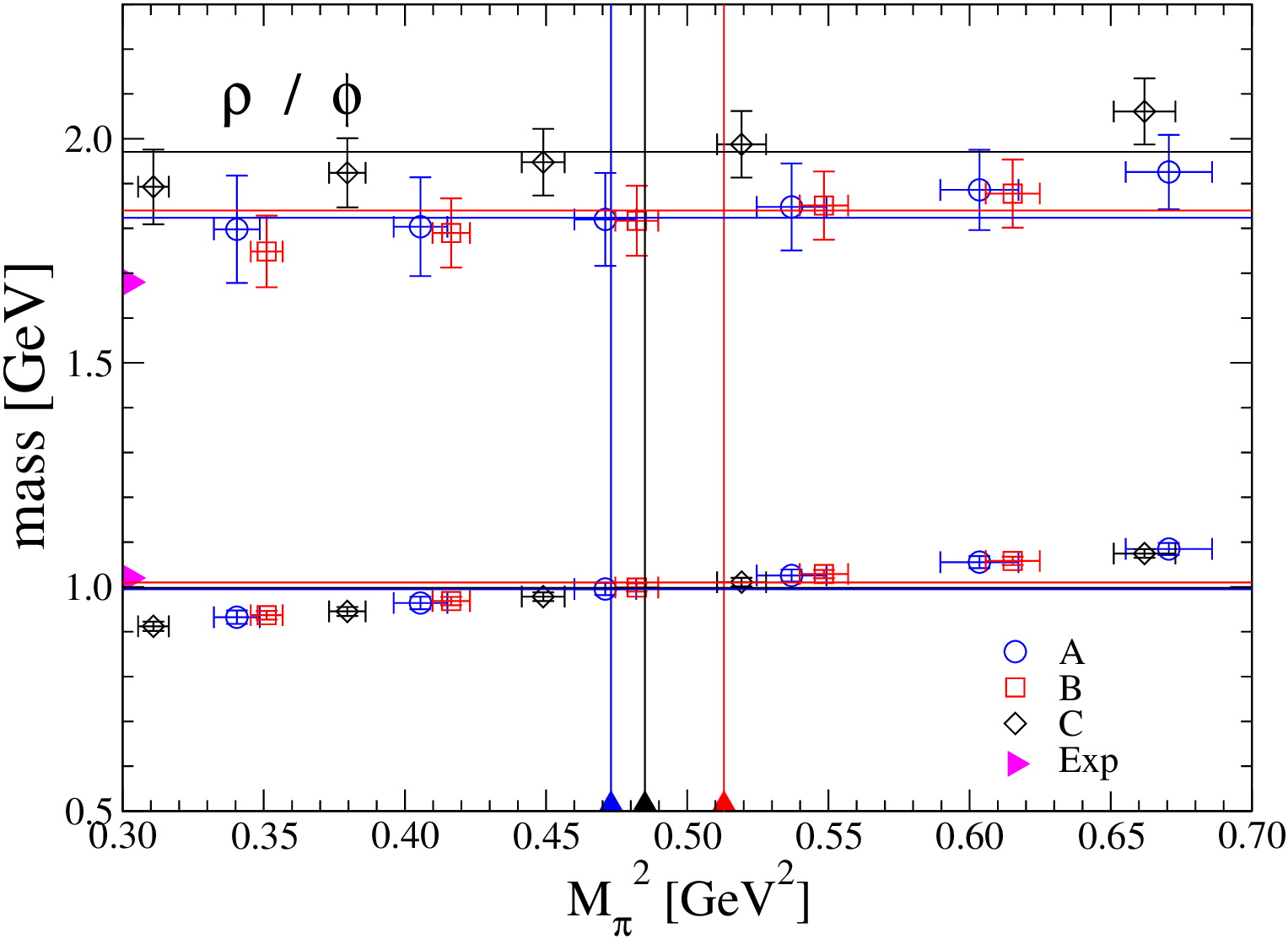}
\caption{Cross-check of the obtained strange quark mass parameter: The partially
quenched $\phi$ from the ground state of the 1$^{--}$ channel fits the
experimental $\phi$(1020) very nicely.  The result for the excited $\phi$ is 
higher than the experimental value, the deviation may be due to the neglected
disconnected diagrams or simply due to the weak signal.}
\label{phi}
\end{minipage}\hfill
\begin{minipage}[t!]{74mm}
\includegraphics[width=74.5mm,clip]{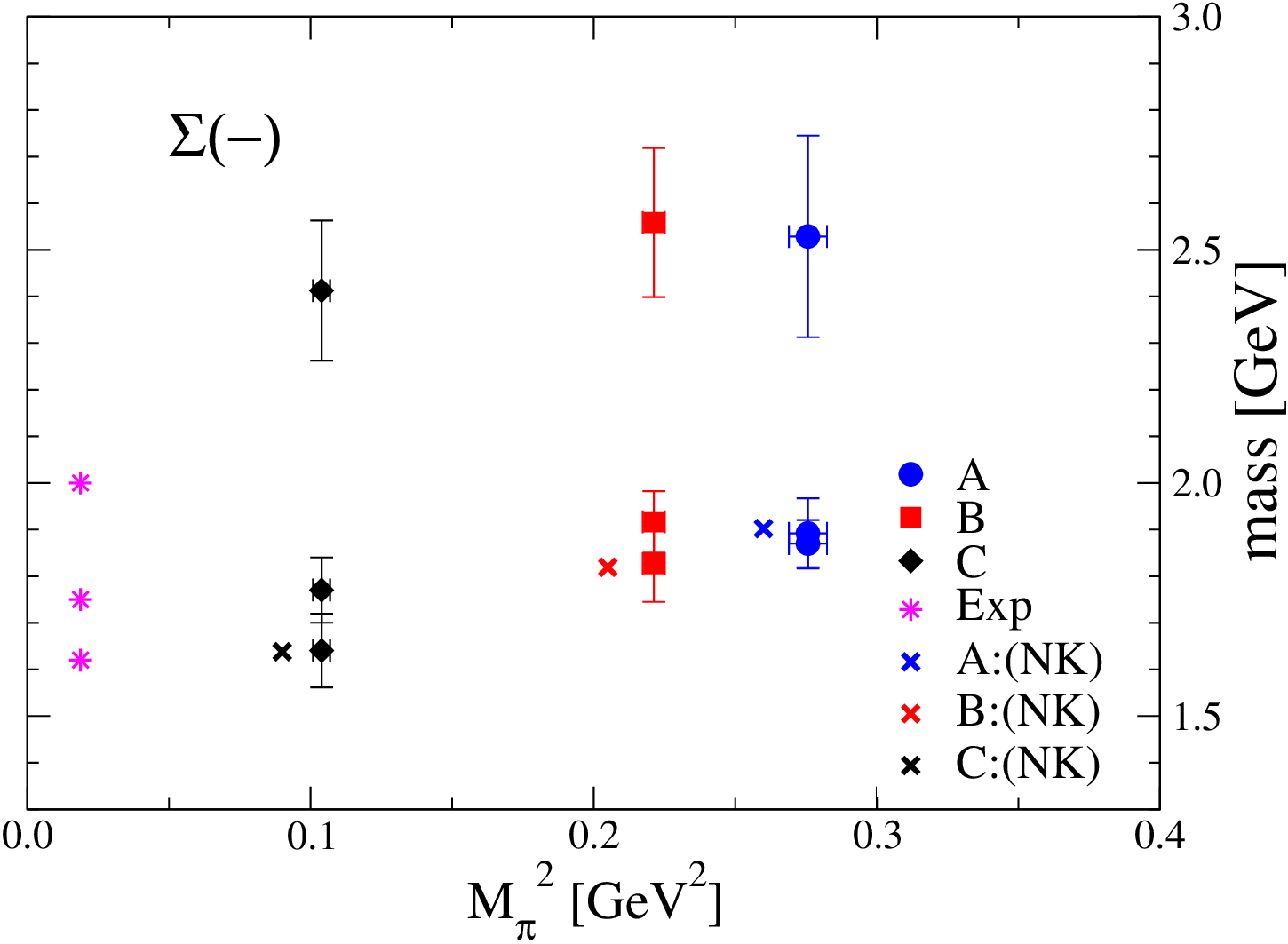}
\caption{Mass plot for the $\Sigma$ negative parity channel (dynamical data only).
For better identification,
we display the scattering states shifted to the left. 
The mass results suggest an interpretation in terms of a level crossing of the 1- and
2-particle ($K N$) states. However, analogously to the nucleon negative parity channel, the
eigenvectors contradict this picture. 
}
\label{sigma_neg}
\end{minipage}
\end{figure}

\noindent
We use our partially quenched results in the $\Delta$ positive parity channel to
identify the strange quark mass parameters % for our ensembles A, B and C 
(see Fig.~\ref{omega}). 
Estimating the mass of the isoscalar $\phi$ from the results for $\rho$ at strange quark mass values serves as a cross-check for the strange quark mass. The result fits the experimental
$\phi$(1020) nicely (see Fig.~\ref{phi}), indicating that our approach is
consistent.  
The ground
state levels of $\Sigma$ and $\Xi$ positive parity provide additional
affirmative cross-checks \cite{Engel:2010my}.

%%%%%%%%%%%%%%%%%%%%%%%%%%%%%%%%%%%%%%%%%%%%%%%%%%%%%%%%%%%%%%%%%%%%%%%%%%%%%%%
\subsection{$\Sigma$ negative parity}
\noindent
In the $\Sigma$ negative parity
channel  we find a ground state and two excitations (see Fig.~\ref{sigma_neg}). Similar to the nucleon negative parity
channel, the results suggest an interpretation in terms of a level crossing
of the 1- and 2-particle ($K N$) states. However, analogously to the nucleon negative parity channel, the
eigenvectors do not support this picture.

%%%%%%%%%%%%%%%%%%%%%%%%%%%%%%%%%%%%%%%%%%%%%%%%%%%%%%%%%%%%%%%%%%%%%%%%%%%%%%%
\section{Conclusions}
\label{Conclusion}
%%%%%%%%%%%%%%%%%%%%%%%%%%%%%%%%%%%%%%%%%%%%%%%%%%%%%%%%%%%%%%%%%%%%%%%%%%%%%%%
%
\begin{figure}[tbp]
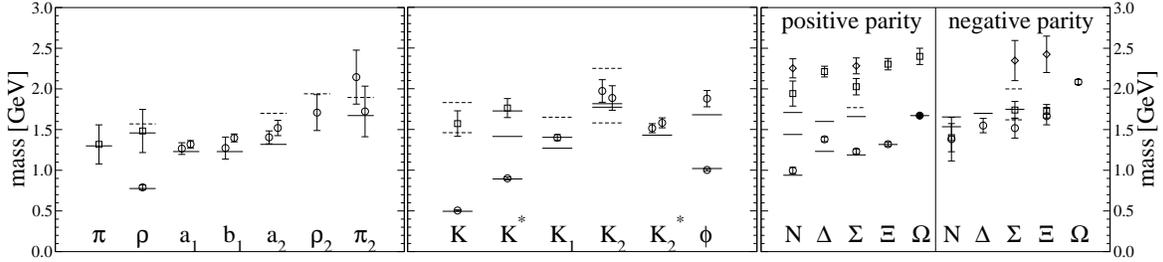

\begin{minipage}[h!]{49mm}
\includegraphics[width=53mm,clip]{./collection_mesons.eps}
\end{minipage}
\hspace{2.2mm}
\begin{minipage}[h!]{46mm}
\includegraphics[width=46.75mm,clip]{./collection_strange_mesons.eps}
\end{minipage}
\begin{minipage}[h!]{49mm}
\includegraphics[width=53mm,clip]{./collection_baryons.eps}
\end{minipage}
\caption{Mass results for light mesons, strange mesons
 and baryons (left to right),  
obtained by chiral extrapolation of dynamical light quarks 
linear in the pion mass squared.
Experimental values listed by the Particle Data Group \cite{PDG08} are denoted
by horizontal lines, the ones needing confirmation by dashed lines.  
Results shown aside each other are obtained using different sets of
interpolators aiming for the same state. 
Strange quarks are implemented by
partial quenching, the strange quark mass parameter is set 
using 
$\Omega$(1670).  Excited baryons 
seem to systematically suffer from finite volume effects. 
Figure taken from \cite{Engel:2010my}.
}
\label{collection_masses}
\end{figure}
\noindent
We presented results of hadron spectroscopy from three ensembles (pion masses from 320 to 530 MeV) using the Chirally Improved 
Dirac operator 
with two light sea quarks (see Fig.~\ref{collection_masses}). 
The strange hadron spectrum was accessed using
partially quenched strange quarks.  
We discussed the possible appearance of scattering states.
The coupling of our interpolators to many-particle  states seems to be weak and
such states are barely, if at all, identifiable. 
Only in the light scalar
channel the partially quenched data suggest a large contribution from an $S$
channel 2-particle state of pseudoscalars. However, at the dynamical point no
clear statement is possible. 
In the negative parity nucleon and $\Sigma$
channels, the eigenvectors do not confirm the picture of the $S$ wave 2-particle
states, either, although such an admixture cannot be completely
excluded. 
Comparison to quenched simulations shows that the spectroscopy in the light scalar and light vector channels seems to benefit slightly from dynamical quarks.
However, in most channels we did not observe a significant difference between
quenched and dynamical simulations.

%%%%%%%%%%%%%%%%%%%%%%%%%%%%%%%%%%%%%%%%%%%%%%%%%%%%%%%%%%%%%%%%%%%%%%%%%%%%%%%
\acknowledgments
%%%%%%%%%%%%%%%%%%%%%%%%%%%%%%%%%%%%%%%%%%%%%%%%%%%%%%%%%%%%%%%%%%%%%%%%%%%%%%%
\noindent
We thank C.R.~Gattringer, L.Y.~Glozman and S.~Prelovsek
for valuable discussions. The calculations have been performed on the SGI Altix
4700 of the Leibniz-Rechenzentrum Munich and on local clusters at ZID at the
University of Graz. M.L.~
and D.M.~have been supported by ``Fonds zur F\"orderung der Wissenschaftlichen
Forschung in \"Osterreich'' (DK W1203	-N08).  D.M.~acknowledges support by
COSY-FFE Projekt 41821486 (COSY-105) and by Natural Sciences and Engineering
Research Council of Canada (NSERC) and G.P.E., M.L.~and A.S.~acknowledge support by
the DFG project SFB/TR-55.

%%%%%%%%%%%%%%%%%%%%%%%%%%%%%%%%%%%%%%%%%%%%%%%%%%%%%%%%%%%%%%%%%%%%%%%%%%%%%%%

%%%%%%%%%%%%%%%%%%%%%%%%%%%%%%%%%%%%%%%%%%%%%%%%%%%%%%%%%%%%%%%%%%%%%%%%%%%%%%%

%%%%%%%%%%%%%%%%%%%%%%%%%%%%%%%%%%%%%%%%%%%%%%%%%%%%%%%%%%%%%%%%%%%%%%%%%%%%%%%
\end{document}